\documentclass[prl,twocolumn,showpacs,amsmath,amssymb,aps]{revtex4}

\usepackage{graphicx}
\usepackage{dcolumn}
\usepackage{bm}

\begin{document}


\title{Thermopower of Aharonov-Bohm Interferometer: Theoretical Studies of
 Quantum Dots in the Kondo Regime}
\author{Tae-Suk Kim$^{1}$ and S. Hershfield$^{2}$}
\affiliation{$^1$ Institute of Physics and Applied Physics, Yonsei University, 
  Seoul 120-749, Korea \\
 $^2$ Department of Physics, University of Florida, Gainesville FL 32611-8440} 
\date{\today}

\begin{abstract}
 We report on the thermopower of an Aharonov-Bohm interferometer (AB)
with a quantum dot in the Kondo regime. 
The thermopower is anomalously enhanced due to the Kondo effect 
as in heavy fermion systems.
In contrast to bulk systems, the sign of the thermopower 
can be changed by adjusting the energy level scheme or
the particle-hole asymmetry of a dot with the gate voltage.
Further the magnitude and even the sign of 
the thermopower in the AB ring
can be changed at will with varying either magnetic fields 
or the gate voltages.

\end{abstract}
\maketitle

 Observations\cite{expkondo1,expkondo2,expkondo3} of the Kondo effect\cite{thekondo} 
in quantum dots have opened a testing ground
of the quantum effects of electron wave functions and many-body effects. 
In contrast to bulk systems where
the phase is washed out by several scattering processes, 
the phase coherence of an electron's wave functions can be preserved in nanoscopic systems. 
A typical system to measure the phase coherence is the Aharonov-Bohm(AB) interferometer. 
Recently a phase-sensitive Fano resonance was observed\cite{fanodotexp} 
in the differential conductance for a quantum dot system which 
contains both a resonant current path and a direct one. 

 An asymmetrical differential conductance suggests that the thermopower
may be the right experimental tool for the study of the phase-sensitive
Aharonov-Bohm interferometer. There are indeed some studies of 
thermopower in AB ring geometries\cite{abring1,abring2} for 
noninteracting electrons. 
The thermopower of electrons is sensitive to the particle-hole
asymmetry in the density of states(DOS) and the energy dependence of 
the electron scattering rate. 
In heavy fermion systems(HFS), the enhanced thermopower and 
its sign are determined by the energy dependence of the Kondo resonance scattering 
rate of the conduction electrons off the magnetic ions\cite{bcwnca,kimcox}. 
In mesoscopic systems, the thermopower will probe sensitively the asymmetrical
shape of the transmission probability.

\begin{figure}[h]
\resizebox{0.5\textwidth}{!}{\includegraphics{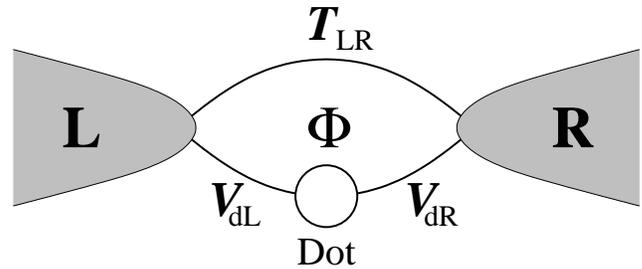}}
\caption{Schematic display of Aharonov-Bohm(AB) interferometer 
with an embedded quantum dot.
The magnetic AB phase $\phi = 2\pi \Phi \cdot e/hc$ is included in the 
tunneling matrices as $V_{dL} T_{LR} V_{Rd} = |V_{dL} T_{LR} V_{Rd}| e^{i\phi}$.
$\Phi$ is the magnetic flux threading through the AB ring. 
\label{fanodot}}
\end{figure}

 In this paper, we study theoretically the thermopower of quantum dot systems
in the Kondo regime. 
The thermopower of a quantum dot is {\it anomalously enhanced} 
due to the Kondo effect as in the HFS. 
In contrast to the HFS, the energy level scheme or the particle-hole 
asymmetry of a dot can be modulated 
continuously by the gate voltage capacitatively coupled to the dot
and the sign of the thermopower can be changed from negative to positive.
When a quantum dot is inserted in the AB ring(see Fig.~\ref{fanodot}), 
the Fano interference\cite{Fano} leads to a more dramatic 
change in thermopower. 
Due to the Kondo effect of a quantum dot, a new Kondo-resonant
current path opens below the Kondo temperature and interferes with the 
direct tunneling path leading to the Fano interference in the 
transmission probability.  
Adjusting both the {\it AB phase} by varying the magnetic field and 
the {\it tunneling matrices} by changing the gate voltages, 
the transmission spectral function 
can be controlled to take several different shapes.
In addition to the differential conductance, the thermopower can probe
the shape of the transmission probability 
since the thermopower is sensitive to 
the particle-hole asymmetry in the transmission probability.  
Combining the Kondo effect and the Fano interference leads to 
an enhanced thermopower of the order of $k_B/e$. 
The magnitude and even the sign of the thermopower can be controlled
with varying the AB phase. 

 We may describe the quantum dot using the Anderson impurity model, 
$H_d =  \epsilon_d \sum_{\alpha=\uparrow,\downarrow} 
      d_{\alpha}^{\dag} d_{\alpha}^{\phantom{*}} 
  + U n_{\uparrow} n_{\downarrow},$
when the number of electrons($N$) in a quantum dot is odd and 
the spin of the highest-lying electron is unpaired.
 Here $\epsilon_d$ is the energy level 
of the highest-lying electron with unpaired spin in a quantum dot
and $U$ is the Coulomb interaction. The left and right leads are 
described by the noninteracting Hamiltonian 
$H_p = \sum_{\vec{k}\alpha} \epsilon_{p\vec{k}} 
    c_{p\vec{k}\alpha}^{\dag} c_{p\vec{k}\alpha}^{\phantom{*}}$
with the lead index $p=L,R$.
The tunneling of electrons between two leads via the direct tunneling($T_{LR}$) 
or through a quantum dot($V_{dp}=V_{pd}^*$) is described by the Hamiltonian,  
$H_1 = [1/V] \sum_{\vec{k}\vec{k}'} \left[ T_{LR} ~
     c_{L\vec{k}\alpha}^{\dag} c_{R\vec{k}'\alpha}^{\phantom{*}} + H.c. \right] 
  + [1/\sqrt{V}] \sum_{p=L,R} \sum_{\vec{k}\alpha} \left[ 
    V_{pd} c_{p\vec{k}\alpha}^{\dag} d_{\alpha}^{\phantom{*}} + H.c. \right].$
The AB phase $\phi = 2\pi \Phi/\Phi_0$ is contained implicitly in the tunneling 
matrices in a manner that $V_{dL} T_{LR} V_{Rd} = |V_{dL} T_{LR} V_{Rd}| e^{i\phi}$. 
$\Phi$ is the magnetic flux passing through the system as shown in 
Fig.~\ref{fanodot} and $\Phi_0 = hc/e$ is the flux quantum. 
Recently, this model system with {\it one conduction channel}
was studied for the conductance  
using the equation of motion method\cite{bulka} 
and the numerical renormalization group in equilibrium\cite{hofstetter}.

 The electric and heat current operators can be defined as a change 
in the number of electrons and the total energy per unit time
in the left electrode, $\hat{I}_L = e[N_L, H]/i\hbar$ and 
$\hat{Q}_L = -[H_L, H]/i\hbar$, respectively.   
Here $N_L = \sum_{\vec{k}\alpha} c_{L\vec{k}\alpha}^{\dag} 
c_{L\vec{k}\alpha}^{\phantom{*}}$ is the number operator and
$H_L$ is the Hamiltonian of the left lead.
We use the Keldysh Green's function method\cite{neqgreen,langreth} 
to write the electric and heat currents. 
Summing over all the multiple tunnelings between two leads and using current 
conservation in a steady state, the electric and heat currents can be expressed 
in terms of the Green's function of a dot\cite{abring1,abring2}. 
\begin{eqnarray}
\label{iqeqn}
\begin{pmatrix} I_L \cr Q_L \end{pmatrix} 
  &=& \frac{2}{h} \int d\omega ~ \begin{pmatrix} -e \cr \omega-\mu_L \end{pmatrix} 
   T(\omega) ~ \left[ f_L (\omega) - f_R (\omega) \right].
\end{eqnarray}
The numerical factor $2$ accounts for the two spin directions and 
$f_p(\omega) = f(\omega-\mu_p)$ is the Fermi-Dirac thermal distribution 
function of $p=L,R$ lead with $\mu_p = -e V_p$.
The {\it transmission spectral function} $T(\omega)$ is given by 
the following expression in the wide conduction band limit
\begin{eqnarray}
\label{tsftn}
T (\omega) 
 &=& T_0 + 2 \overline{\Gamma} \sqrt{g T_0 (1-T_0)} ~\cos\phi ~\mbox{Re} G_d^r 
    \nonumber\\
 && + \overline{\Gamma} \left[ T_0 - g (1 - T_0\cos^2\phi) \right]~
     \mbox{Im} G_d^r.     
\end{eqnarray} 
Here $G_d^{r}$ is the retarded Green's function of a quantum dot and
$\overline{\Gamma} = (\Gamma_L + \Gamma_R)/(1+\gamma)$ is the Anderson
hybridization. 
$\Gamma_p = \pi N_p |V_{dp}|^2$ 
measures the hopping rate of electrons between the quantum dot and the leads, 
where $N_p$ is the density of states of lead $p=L,R$.
$T_0=4\gamma /(1+\gamma)^2$ is the transmission probability 
due to the direct tunneling, and 
$\gamma = \pi^2 N_L N_R |T_{LR}|^2$ is the dimensionless measure of direct 
tunneling of electrons between the two leads. 
$g=4\Gamma_L \Gamma_R /( \Gamma_L + \Gamma_R)^2$ 
is the maximum dimensionless linear conductance through a quantum dot in the
absence of the direct tunneling and also measures asymmetry in the coupling of 
a quantum dot to the left and right reservoirs. 
The interference effect is included
in the second and third terms of $T (\omega)$.
The above expression for $T(\omega)$ agrees with Eq. (2) 
in Ref. \cite{hofstetter}.

 We compute the dot's Green's function $G_d$ using the noncrossing 
approximation(NCA) for the infinite $U$ Anderson model in the Kondo regime.
The NCA\cite{eqnca1,eqnca2,neqnca1,neqnca2,kimselman}
 has been successfully used for the study of 
the Anderson model except for the nonanalytic behavior\cite{zeronca} 
at a temperature far below the Kondo temperature $T_K$.\cite{nonanal}  
In the NCA self-energy equations, the renormalized Anderson hybridization 
should be used instead of the bare Anderson hybridization\cite{tobe}.
The multiple tunnelings between the two leads result in the flux-dependent 
renormalized Anderson hybridizations
which in a wide conduction band limit are given by the equations,
\begin{eqnarray}
\label{ahyb}
\overline{\Gamma}_{L,R}
 &=& \frac{1}{(1 + \gamma)^2} \left[ \Gamma_{L,R} + \gamma \Gamma_{R,L} 
    \mp 2\sqrt{\gamma \Gamma_L \Gamma_R} \sin \phi \right]. 
\end{eqnarray}
In equilibrium, the two thermal functions are equivalent or $f_L = f_R$ so that 
the total Anderson hybridization 
$\overline{\Gamma} = \overline{\Gamma}_{L} + \overline{\Gamma}_{R}$ 
is independent of the AB phase $\phi$ and so is $G_d^{r}$.
$T(\omega)$ also remains invariant under the inversion 
of the magnetic flux: $\Phi \to -\Phi$. 
The Onsager relation $T(\omega,-\phi) = T(\omega,\phi)$ in equilibrium 
becomes broken when a finite source-drain bias voltage is applied
($f_L \neq f_R$). 
The Kondo temperature $T_K$ is independent of $\phi$ and 
can be estimated in the $U\to\infty$ limit by 
the equation $T_K = D \sqrt{N(0)J} ~\exp\left(-{1/N(0)J} \right)$
with $N(0)J = 2\overline{\Gamma} /\pi |\epsilon_d|$. 
Note that the direct tunneling suppresses the Kondo effect.

 At high temperature above $T_K$, the current flow through a quantum dot 
is blocked due to the strong Coulomb repulsion (Coulomb blockade). 
Electrons flow from the left reservoir to the right one 
only via the direct tunneling. With decreasing temperature below $T_K$,
the Kondo resonance peak at the quantum dot develops close to the Fermi level. 
The newly opened current path interferes with the direct tunneling path. 
This Fano interference transforms $T(\omega)$ into various different shapes
depending on the AB phase $\phi$. 
 The general structure of $T(\omega)$ near $\omega=0$ 
can be read off from the Eq. (\ref{tsftn}). It is well known that 
$-\mbox{Im} G_d^{r}$ develops the Kondo resonance peak with its width 
of the order of $T_K$ near $\omega=0$ while $\mbox{Re} G_d^{r}$ 
varies very rapidly over the energy scale of $T_K$ near $\omega=0$ with 
 a dip just below $\omega = 0$ and a peak above $\omega=0$\cite{tobe}. 
The overall shape of the transmission spectral function $T(\omega)$ 
is determined by the value of the AB phase $\phi$ and 
the sign of $\Delta_c$[see Eq.(\ref{tsftn})], 
\begin{eqnarray}
\Delta_c &\equiv& T_0 - g(1-T_0\cos^2\phi).
\end{eqnarray}  
A typical Fano interference pattern consisting of a dip and peak
structure is expected when $\cos\phi \neq 0$. 
At $\cos\phi = 0$, $T(\omega)$ has a dip(peak) resonance structure
if $\Delta_c >(<) 0$, respectively.

 In our numerical NCA work we consider a symmetrically coupled dot
($g=1$ or $\Gamma_L=\Gamma_R$) with the energy level scheme(ELS): 
the $N-1$ state lies lower in energy than the $N+1$ state,
$E_{N-1} \ll E_{N+1}$,  
where $N$ is the number of electrons in a dot. 
Other ELS's will be discussed later. 
In the Anderson model picture, $E_{N-1} =0$ (empty), $E_N = \epsilon_d$
(singly occupied), and $E_{N+1} = 2\epsilon_d + U$ (doubly occupied). 
In practice, we take the limit of $U\to\infty$\cite{grewe} and
the DOS of two leads is assumed to be Lorentzian of bandwidth $D$.  
%

\begin{figure}
\noindent
\begin{minipage}[t]{.85\linewidth}
 \includegraphics{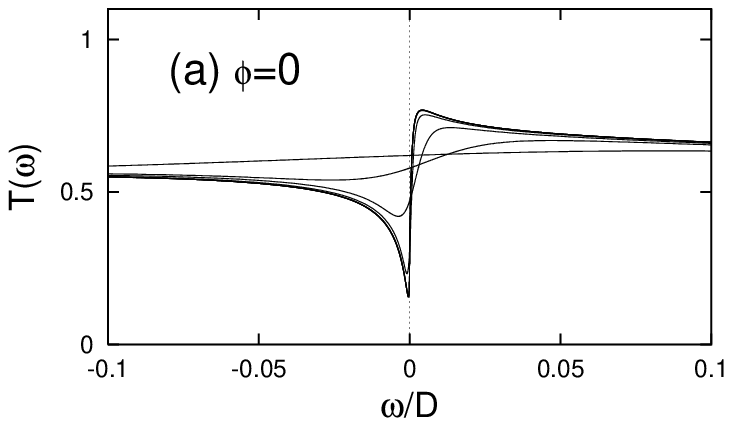}
 \includegraphics{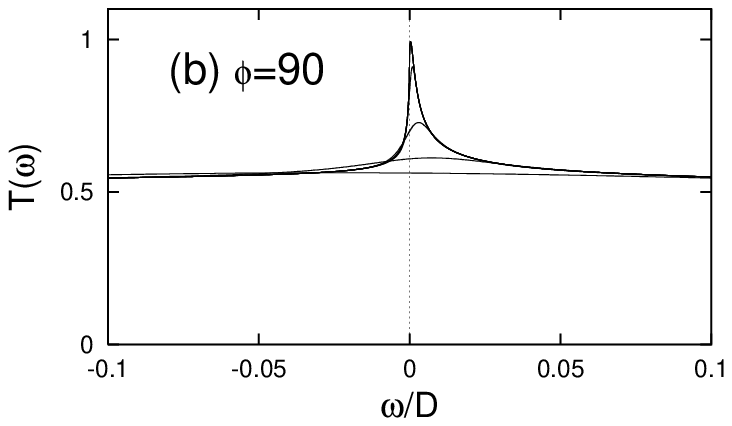}
 \includegraphics{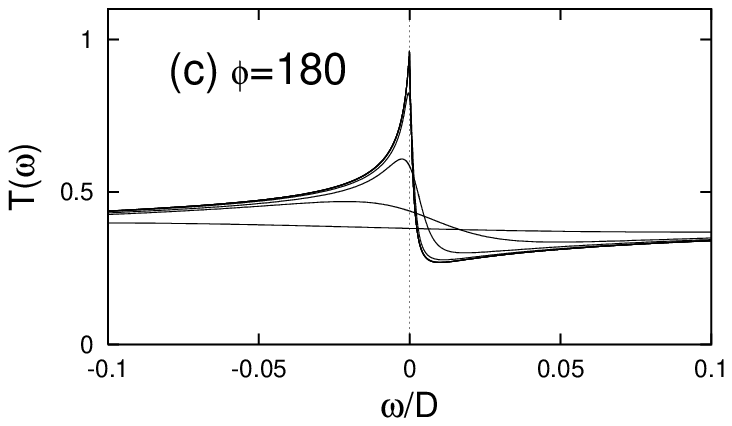}
\end{minipage}
\caption{Dependence of the transmission probability $T(\omega)$ near the Fermi level 
on temperature $T$, the AB phase $\phi$. 
The model parameters are chosen as $T_0=0.5$, $\epsilon_d/D = -0.5$, 
and $\overline{\Gamma}_L=\overline{\Gamma}_R=0.07 D$. 
With lowering T, the Kondo correlation sharpens the shape
of $T(\omega)$. $T$ is varied as 
$T/T_K = 100, 20, 4, 0.8, 0.16, 3.2\times 10^{-2}, 6.3\times 10^{-3},
1.3\times 10^{-3}$.  
The last four temperature curves cannot be distinguished with the naked eye.
Fano interference between the direct path 
and the Kondo-resonant tunneling leads to the strong dependence of 
$T(\omega)$ near the Fermi level on the AB phase.    
\label{fanos}}
\end{figure}

For the weak direct tunneling, the Fano interference remains weak and 
the Kondo-related peak persists in $T(\omega)$ over all $\phi$\cite{tobe}.
The AB phase $\phi$ dependence of $T(\omega)$ becomes stronger 
with increasing direct tunneling amplitude $T_0$. 
The case of $T_0=0.5$ is displayed in Fig.~\ref{fanos}
and the shape of $T(\omega)$ near $\omega=0$ is strongly sensitive to 
the AB phase $\phi$.
Since $\Delta_c = 0$ at $\phi=0^{\circ}$ and $180^{\circ}$, 
$T(\omega)$ has the dip-and-peak structure of 
$\mbox{Re} G_d^r$ (see Fig.~\ref{fanos}(a) and (c)).

When $\cos \phi = 0$ ($\phi=90^{\circ}$,$270^{\circ}$),
the spectral shape of $T(\omega)$ is wholly determined by 
both $\mbox{Im} G_d^r$ and the sign of $\Delta_c$.
$\Delta_c$ remains always negative for $g=1$
so that the Kondo resonance peak persists in $T(\omega)$ for the whole range of 
$T_0 (0\leq T_0 \leq 1)$. 
The Fermi-liquid relation\cite{langrethfl}, 
$\mbox{Im} G_d^r(0) = -1/\overline{\Gamma}$ at $T=0$K, leads to 
one {\it exact} relation:  
$T(0) = g$ for any interacting dots.
$T(\omega)$ reaches its maximum possible value
$g=1$ at $\omega=0$, a unitary Kondo resonance tunneling. 
When the background direct tunneling amplitude $T_0$ is increased from $0$ to $1$, 
$|\Delta_c|$ approaches zero and the Kondo-related peak becomes smaller\cite{tobe}.

 Except for close to $\phi=90^{\circ}$, $T(\omega)$ of
asymmetrically coupled dots($g<1$) is not much different from the symmetrically 
coupled dots\cite{tobe}. 
Since $\Delta_c=T_0-g$ at $\phi=90^{\circ}$ can change its sign,
the Kondo-related peak in $T(\omega)$ transforms into a dip
at $\omega=0$ as the value of $T_0$ crosses $g$. 
$T(\omega)$ is flat near the Fermi level when $T_0=g$\cite{tobe}.

 The thermopower of a quantum dot 
in a two-terminal configuration
can be found in an open circuit($I=0$) by measuring the induced voltage
drop across a quantum dot when a temperature difference between the two leads
is applied. The thermopower $S$ is defined by the relation,
\begin{eqnarray}
S &\equiv& - \lim_{T_L \to T_R} \left. \frac{V_L - V_R}{T_L - T_R} \right|_{I=0}. 
\end{eqnarray} 
Expanding the expressions for $I$ and $Q$ in Eq.(\ref{iqeqn})
up to the linear terms of $\delta V = V_L - V_R$ and
$\delta T = T_L - T_R$, the transport coefficients can be 
expressed in terms of the integral, 
$I_n (T) \equiv [{2 / h}] \int d\omega ~ \omega^n T(\omega) 
   \left[ -{\partial f / \partial \omega} \right]$. 
$I = L_{11} \delta V + L_{12} \delta T$ and 
$Q = L_{21} \delta V + L_{22} \delta T$ where 
$L_{11} = e^2 I_0(T)$, 
$L_{21} = L_{12} T = - e I_1(T)$, and 
$L_{22} = I_2(T)/T$.

\begin{figure}
\includegraphics{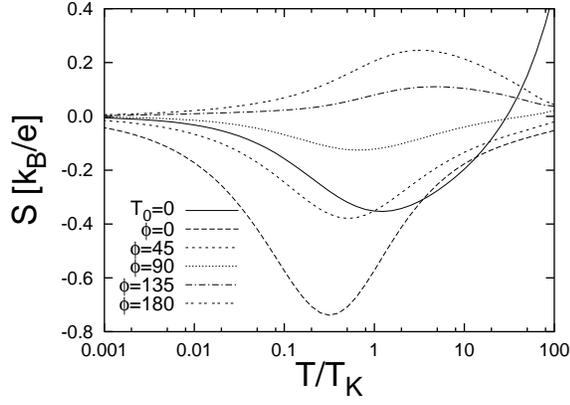} 
\caption{Thermopower $S$. 
(i) The solid line is the thermopower
when the direct tunneling is absent or $T_0=0$. Due to the Kondo effect,
$S$ is enhanced near the Kondo temperature and negative in the energy level
scheme $E_{N-1} \ll E_{N+1}$. 
(ii) The five broken lines show the dependence of $S$ on the AB phase $\phi$. 
Due to Fano interference,
the magnitude and sign of $S$ are modulated by the magnetic AB flux.    
\label{tpower}}
\end{figure}

 The thermopower $S = - I_1 / e T I_0$ probes  
the particle-hole asymmetrical part of $T(\omega)$.
To begin we compute the thermopower in the absence of direct 
tunneling or $T_0=0$. In our choice of the ELS in a dot,
$E_{N-1} \ll E_{N+1}$, the Kondo resonance peak has more spectral weight 
on the electron excitations. Since the electron excitations are the main 
carriers of charge and heat, the sign of the thermopower is negative. 
From the study of the heavy fermion systems(HFS), it is well known that
the thermopower is anomalously enhanced due to the Kondo effect
and is of the order of $k_B/e (\approx 86.17\mu$V/K) near $T=T_K$
\cite{bcwnca,kimcox}. 
In normal metals, the thermopower is of order $\mu$V/K. 
The thermopower is also enhanced in a quantum dot
in the Kondo regime, as shown in Fig.~\ref{tpower} by the solid line.

 To study the effect of the Fano interference on $S$, we now turn on the 
direct tunneling. 
The computed thermopower $S(T)$ for $T_0 = 0.5$ is displayed 
in Fig.~\ref{tpower} for different AB phase $\phi$. 
When $\phi = 0^{\circ}$, the electron transmission is 
high while the hole transmission is low[Fig.~\ref{fanos}(a)]. 
Since electron excitations are the main carriers of charge and heat, 
the thermopower is negative
and its magnitude is of the order of $k_B/e$ near $T=T_K$ 
due to the Kondo effect.
With increasing AB phase, the dip-peak structure in $T(\omega)$ transforms into 
a Kondo-related peak at $\phi=90^{\circ}$[Fig.~\ref{fanos}(b)]. 
In this case, $T(\omega)$ is 
more or less symmetrical with respect to $\omega=0$ though more spectral 
weight lies on the electron excitations.
$S$ is therefore weakly negative over a large temperature range. 
When $\phi = 180^{\circ}$, the hole (electron) transmission 
is high (low)[Fig.~\ref{fanos}(c)]. 
Since the holes are the main carriers, the thermopower is positive. 
In summary when the quantum dot is inserted in the AB interferometer, 
the magnitude and the sign of $S$ 
can be changed by varying the AB phase $\phi$ or magnetic fields threading
the AB ring.

 We now address the effect of the energy level scheme(ELS) in a quantum dot
on the thermopower when $T_0=0$. 
 The sign and the magnitude of the thermopower can be controlled by 
adjusting the ELS in a dot, too. 
The results in Fig.~\ref{tpower} are 
computed for the dot with $E_{N-1} \ll E_{N+1} (U\to \infty)$.
In analogy to the heavy fermion systems(HFS), this ELS 
is equivalent to the Ce alloys where the doubly occupied 
$f$-electron orbitals lie well above the empty $f$-electron orbital state. 
Adjusting the gate voltage capacitatively coupled to the dot, 
the ELS of a dot can be changed in a continuous manner. 
When $E_{N-1} > E_{N+1}$, 
more spectral weight of the Kondo resonance peak lies on the 
hole excitations. 
The main carriers are hole excitations, 
leading to a positive thermopower. 
This inverted ELS is a particle-hole symmetric image of an ELS
$E_{N-1} < E_{N+1}$ with respect to the point $E_{N-1} = E_{N+1}$
and corresponds to the Yb alloys in HFS.
Since the flow of electrons is {\it Kondo-assisted} in a quantum dot, 
the sign of $S$ is negative(positive) when $E_{N-1} <(>) E_{N+1}$, 
respectively. In the Ce or Yb HFS, electrons are {\it Kondo-scattered}
so that the sign of $S$ is positive(negative) for the Ce(Yb) alloys,
respectively. Note that the sign of the thermopower is opposite 
in a quantum dot and the bulk HFS with the same energy level scheme. 
When the ELS in a dot is particle-hole symmetric($E_{N-1} = E_{N+1}$), 
the Kondo resonance peak of the dot is also symmetric with respect to the 
Fermi level. 
Since the heat currents carried by electrons and holes are canceled 
by each other, the resulting thermopower is zero. 
The linear conductance measures the symmetrical part of $T(\omega)$ 
near $|\omega| < k_BT$ and its value is not sensitive to the 
shape of $T(\omega)$. On the other hand, the sign of the thermopower is
sensitive to the degree of the particle-hole asymmetry in $T(\omega)$. 
Hence the sign of $S$ can probe the energy level scheme in a quantum dot.

 In summary, we studied the thermopower in a quantum dot in the Kondo regime.
In contrast to the bulk heavy fermion systems, the sign of the thermopower 
in a quantum dot can be changed by adjusting the energy level scheme
or the particle-hole asymmetry in a dot. When the dot is inserted in an
AB interferometer, the dramatic variations in the shape of the transmission 
spectral function $T(\omega)$ are possible by controlling 
the AB phase and the tunneling matrices. 
The rich asymmetrical shapes of $T(\omega)$ manifest themselves in the 
magnitude and the sign of the thermopower. 
Combination of the Kondo effect and the Fano interference
enables us to control the magnitude and the sign of thermopower.

\acknowledgments
This work was supported in part by the National 
Science Foundation under Grant No. DMR 9357474, 
in part by the BK21 project and
in part by grant No. 1999-2-114-005-5 from the KOSEF.

%
%
%

%
%


\begin{thebibliography}{100}
\bibitem{nonanal} In our work, the temperature below which the nonanalyticity 
shows up is estimated to be less than $10^{-4} T_K$.
\end{thebibliography}
\end{document}